 \documentclass[final,5p,times,twocolumn]{elsarticle}

\usepackage{amssymb,amsmath}
\usepackage{lipsum}
\usepackage[colorlinks,linkcolor=blue,anchorcolor=blue,citecolor=blue]{hyperref}

\journal{Physics Letters B}

\begin{document}

\begin{frontmatter}



\title{\textit{Ab initio} study in the island of inversion within the two-major-shell valence space}


\author[sysu,klqms]{X.~C.~Cao}
\author[sysu,klqms]{C.~F.~Jiao\corref{cor1}}
\ead{jiaochf@mail.sysu.edu.cn}
\cortext[cor1]{Corresponding author.}
\affiliation[sysu]{organization={School of Physics and Astronomy},
            addressline={Sun Yat-sen University}, 
            city={Zhuhai},
            postcode={519082}, 
            state={Guangdong},
            country={China}}
\affiliation[klqms]{organization={Guangdong Provincial Key Laboratory of Quantum Metrology and Sensing},
            addressline={Sun Yat-sen University}, 
            city={Zhuhai},
            postcode={519082}, 
            state={Guangdong},
            country={China}}

\begin{abstract}
We present an \textit{ab initio} study of nuclear structure in the island of inversion around neutron number $N=20$, using multishell effective Hamiltonians derived from the valence-space in-medium similarity renormalization group approach combined with the quantum-number projected generator coordinate method. By progressively expanding the valence space from the \textit{sd} shell to the intermediate $sdf_{7/2}p_{3/2}$ space and, for the first time, to the full \textit{sdfp} shell, we investigate low-lying spectra, $E2$ transition strengths, deformation properties, and neutron occupancies in even-even Ne, Mg, and Si isotopes around $N=20$. Our results show that enlarging the valence space significantly improves the description of quadrupole collectivity, yielding better agreement with experimental data for key observables such as the lowered $2^+$ energies and the enhanced $B(E2;0^+_1\rightarrow2_1^+)$ values. The analysis reveals the critical role of cross-shell multi-particle multi-hole excitations in breaking the $N=20$ shell closure and establishing intruder-dominated ground states. It also demonstrates the ability of the VS-IMSRG+PGCM framework to capture both dynamical (short range) and static (long range) correlations across multiple major-oscillator shells.
\end{abstract}




\end{frontmatter}




\section{Introduction}
\label{introduction}

Shell structure is a common characteristic of a finite quantum many-body system. In atomic nuclei, the shell structure and the associated ``magic numbers'', governed by inter-nucleon interactions, have been extensively studied since the pioneering work of Mayer and Jensen~\cite{MayerPR1949,HaxelPR1949}. The advent of large-scale rare-isotope beam facilities has opened up opportunities to explore novel shell structures in unstable exotic nuclei. A particularly intriguing discovery is the so-called \textit{island of inversion}~\cite{PhysRevC.90.014302,PhysRevC.95.021304,RevModPhys.77.427,RevModPhys.92.015002,Tsunoda2020}, where conventional shell closures break down and the nuclei exhibit unexpectedly enhanced quadrupole collectivity~\cite{PhysRevC.102.034320,PhysRevC.41.1147,PhysRevC.90.014302}. This phenomenon was first identified through the disappearance of the neutron magic number $N = 20$ in neutron-rich neon, sodium, and magnesium isotopes~\cite{MOTOBAYASHI19959,PhysRevC.41.1147,PhysRevC.12.644}, challenging the traditional understanding of magic numbers and sparking considerable interest in elucidating the mechanisms underlying shell evolution~\cite{PhysRevLett.95.232502,PhysRevC.60.054315,POVES1987311}. 

Substantial efforts, both experimental~\cite{PhysRevC.12.644,PhysRevC.19.164,GUILLEMAUDMUELLER198437,MOTOBAYASHI19959,IWASAKI2001227,PhysRevLett.99.212501,PhysRevLett.105.252501,PhysRevLett.111.212502,PhysRevC.93.044306} and theoretical~\cite{CAMPI1975193,POVES1987311,PhysRevC.58.2033,PhysRevC.60.054315,PhysRevC.69.034301,PhysRevC.95.021304}, have been devoted to  understanding the island of inversion.
These studies have established that the ground-state (g.s.) wave functions of nuclei in this region are dominated by intruder configurations, primarily involving multi-particle multi-hole (\textit{mp-mh}) excitations from the $sd$ to the $fp$ orbits~\cite{PhysRevC.90.014302,ZHOU2025139464,PhysRevC.102.034320}. This is attributed to a weakening of the $N = 20$ shell gap between the two major oscillator shells, leading to the disappearance of the traditional magic number. %
Although certain phenomenological approaches have succeeded in reproducing the enhanced quadrupole collectivity, these studies typically rely on nuclear interactions that are empirically fine-tuned to existing data on finite nuclei. This dependence limits their predictive power for exotic nuclei far from the valley of stability, where experimental data are scarce. 

During the past two decades, significant progress has been made in developing \textit{ab initio} nuclear structure methods~\cite{LEE2009117,Navrátil_2009,BARRETT2013131,PhysRevC.89.061301,Hagen_2014,RevModPhys.87.1067,LAUNEY2016101,HERGERT2016165}. These approaches aim to describe nuclear properties directly from the fundamental inter-nucleon interactions [e.g., nucleon–nucleon (\textit{NN}) and three-nucleon (3\textit{N}) interactions derived from chiral effective field theory ($\chi$EFT)] and retain the connection to quantum chromodynamics (QCD)~\cite{WEINBERG19913,MACHLEIDT20111,RevModPhys.81.1773}, without relying on empirical adjustments beyond those necessary to constrain the nuclear forces themselves. Despite promising advances in \textit{ab initio} methods, the microscopic understanding of the island of inversion from the fundamental nuclear forces still poses a challenge. Recently, the extended Kuo-Krenciglowa (EKK) method based on many-body perturbation theory (MBPT) has been used to perturbatively derive an effective Hamiltonian within the \textit{sdfp} shell from the chiral nuclear force~\cite{PhysRevC.95.021304}, but it still empirically fits the single-particle energies to accurately describe the island of inversion.

The valence-space in-medium similarity renormalization group (VS-IMSRG) is a powerful non-perturbative \textit{ab initio} framework capable of computing a wide range of nuclear observables—including energies, charge radii, and electroweak moments and transitions—for both ground and excited states in open-shell nuclei~\cite{PhysRevLett.118.032502,PhysRevLett.113.142501,PhysRevC.93.051301,PhysRevC.96.034324,annurev:/content/journals/10.1146/annurev-nucl-101917-021120,PhysRevC.102.034320}. It starts from chiral \textit{NN} and 3\textit{N} interactions and non-perturbatively decouples an effective Hamiltonian within a chosen valence space. The diagonalization of this Hamiltonian, typically via the shell model, then yields the eigenvalues and wave functions of the many-body system. However, a major difficulty in applying the VS-IMSRG to the island of inversion is the need to incorporate intruder configurations that involve multiple major shells (the \textit{sd} and \textit{fp} shells). In conflict with it, deriving effective Hamiltonians has generally been restricted to single major-shell valence spaces due to the computational limitations of conventional shell model diagonalization. Recently, the VS-IMSRG derived the first \textit{ab initio} multishell valence-space Hamiltonians for this region~\cite{PhysRevC.102.034320,PhysRevC.111.054308}, but the valence space is limited to the $\pi$\textit{sd}+$\nu$\textit{sd}$f_{7/2}p_{3/2}$ shell. It does not reproduce the lowered 2$_1^+$ state and strong electric quadrupole transitions for $^{32}$Mg, indicating that the effective Hamiltonians within the $\pi$\textit{sd}+$\nu$\textit{sd}$f_{7/2}p_{3/2}$ shell may not fully capture the enhanced quadrupole collectivity. A complete treatment of cross-shell \textit{mp-mh} excitations from the \textit{sd} to the \textit{fp} shells requires effective Hamiltonians spanning the full two major oscillator shells (\textit{sdfp}).

Diagonalizing \textit{sdfp}-shell Hamiltonians is computationally prohibitive within the conventional shell model. Alternatively, the quantum-number projected generator coordinate method (PGCM) has been established as an outstanding variational approximation to exact diagonalizations in shell-model valence spaces~\cite{PhysRevC.96.054310,PhysRevC.98.064324,PhysRevC.98.054311,PhysRevC.100.044308,PhysRevC.100.031303,PhysRevC.103.064302,PhysRevC.110.054326}. In addition, the PGCM can be readily extended to very large model spaces~\cite{PhysRevC.96.054310,PhysRevC.110.054326}. Similarly to the spirit of PGCM with explicit quasiparticle excitations, the discrete non-orthogonal shell model (DNO-SM) has been benchmarked against a full set of \textit{sd} shell exact diagonalizations and applied to heavy deformed nobelium nuclei~\cite{PhysRevC.105.054314,arXiv:2409.08210}.
Given the demonstrated success of the valence-space extension within the VS-IMSRG framework in improving the descriptions of cross-shell correlations and enhanced quadrupole collectivity~\cite{PhysRevC.102.034320,PhysRevC.111.054308}, the PGCM presents a natural choice over the conventional shell model. It should be noted that a different theoretical approach that combined the density-matrix renormalization group (DMRG) with the VS-IMSRG also leads to an efficient hybrid solver for nuclear many-body problems~\cite{TICHAI2023138139,TICHAI2024138841}.

In this work, we exploit the VS-IMSRG approach to decouple the \textit{ab initio} valence-space Hamiltonians spanning the \textit{sd}, the $sdf_{7/2}p_{3/2}$, and the \textit{sdfp} shell. These effective Hamiltonians are then used in PGCM calculations to systematically study the island of inversion with progressively expanded valence spaces. We demonstrate the improvement achieved by the expansion of the valence spaces in describing the low-lying spectroscopic properties of island-of-inversion nuclei using the VS-IMSRG effective Hamiltonians. We also provide a detailed analysis of the \textit{mp-mh} excitations from the \textit{sd} to the \textit{fp} shell across two full major oscillator shells. Other novel \textit{ab initio} frameworks, such as the angular-momentum projected coupled cluster theory with singles and doubles (AMP-CCSD)~\cite{PhysRevC.105.064311,PhysRevX.15.011028,PhysRevC.111.044304} and the in-medium generator-coordinate method (IM-GCM)~\cite{PhysRevC.98.054311,PhysRevLett.124.232501} that combines the PGCM and multi-reference IMSRG (MR-IMSRG)~\cite{PhysRevLett.110.242501, 10.3389/fphy.2020.00379}, have also been applied to this region. Although achieving remarkable success in describing strongly deformed rotational bands~\cite{PhysRevC.105.064311,PhysRevX.15.011028,PhysRevC.111.044304} and shape co-existence~\cite{ZHOU2025139464}, their applications to heavier nuclei are severely constrained by high computational costs. Alternatively, the valence-space-based VS-IMSRG+PGCM framework employed here offers a scalable path for future \textit{ab initio} studies of heavier open-shell nuclei.

\section{The model}
In the valence-space framework, the single-particle Hilbert space is partitioned into core, valence, and outside subspaces.
The core orbitals (i.e., $^{16}$O in this work) and outside orbitals are treated as inactive in the final calculations.
In the valence space, the effective Hamiltonian, which encodes the essential degrees of freedom for reproducing low-lying states, is constructed to account for possible configuration mixing through exact diagonalization or the PGCM approach used in the present work. Therefore, our objective is to develop an effective Hamiltonian where excitations out of the core or into the outside subspaces are rigorously decoupled. To achieve this decoupling, the IMSRG evolves the initial Hamiltonian via a continuous unitary transformation governed by the flow equation given by

\begin{equation}
  \frac{dH(s)}{ds} = [\eta(s),H(s)],
\end{equation}
where the Hamiltonian, which depends on the flow parameter $s$, is expressed in the second-quantized form:
\begin{equation}
  H(s) = E_{0} + \sum_{ab}f_{ab}(s)\{a_{a}^\dagger a_{b}\} + \frac{1}{4}\sum_{abcd}\Gamma_{abcd}(s)\{a_{a}^\dagger a_{b}^\dagger a_{d} a_{c}\}.
\end{equation}
$E_0$, $f_{ab}$, and $\Gamma_{abcd}$ denote the zero-, one-, and two-body matrix elements of the Hamiltonian, respectively. %
$a_a$ ($a_a^\dagger$) is the operator that annihilates (creates) a particle in the single-particle orbital labeled by $a$. %
The $\{ \cdots \}$ indicates normal ordering with respect to the reference state, which can be a single Slater determinant or an ensemble state~\cite{PhysRevLett.118.032502}. %
The anti-Hermitian operator $\eta(s)$ is the so-called generator written as
\begin{equation}
  \eta = \sum_{ai}\eta_{ai} \{a_{a}^\dagger a_{i}\} + \sum_{abij}\eta_{abij} \{a_{a}^\dagger a_{b}^\dagger a_{j} a_{i}\} -\text{H.c}.
\end{equation}
with 
\begin{equation}
  ai \in \{pc,ov\},  \quad abij \in \{pp'cc',pp'vc,opvv'\}.
\end{equation}
The indices $c, v$, and $o$ indicate core, valence, and outside space orbitals, respectively, and $p$ denotes either $v$ or $o$. Because we need to derive effective Hamiltonians spanning multiple major oscillator shells, the generator is defined as
\begin{eqnarray}
  \eta_{ai} &=& \frac{1}{2}\arctan \left(\frac{2f_{ai}}{f_{aa}-f_{ii}+\Gamma_{aiai}+\Delta}\right), \label{generator1}\\ 
  \eta_{abij} &=& \frac{1}{2}\arctan \left(\frac{2\Gamma_{abij}}{f_{aa}+f_{bb}-f_{ii}-f_{jj}+G_{abij}+\Delta}\right), \label{generator2}\\
  G_{abij} &=& \Gamma_{abab}+\Gamma_{ijij}-\left(\Gamma_{aiai}+\Gamma_{bjbj}+[a \leftrightarrow b]\right)\label{generator3}.
\end{eqnarray}
Here, $\Delta$ denotes the energy denominator shift which is introduced to address a known issue in the decoupling of multi-shell Hamiltonians~\cite{PhysRevC.102.034320}. The issue is that, as the flow parameter $s$ increases, some single-particle levels in the outside space may evolve downward and drop below the valence-space levels~\cite{PhysRevC.93.051301,PhysRevLett.118.032502}. A detailed numerical analysis of the effect of the energy denominator shift $\Delta$ can be found in Ref.~\cite{PhysRevC.102.034320}.\par 

It should be noted that the IMSRG evolution induces three- and higher-body terms, which should be kept in principle, but are computationally impractical. Here we keep up to
two-body terms (known as the IMSRG(2) approximation), which
has been proven to be an effective many-body truncation when applied to nuclei in different mass regions~\cite{HERGERT2016165,PhysRevC.85.061304,PhysRevLett.106.222502,PhysRevLett.110.242501,PhysRevLett.120.152503}.

As the effective Hamiltonian is obtained in a large model space where conventional shell-model calculations become computationally intractable, we instead exploit the PGCM based on the evolved Hamiltonian $H(s)$. It has been proven to provide a great variational approximation to the conventional shell model~\cite{PhysRevC.96.054310,PhysRevC.100.044308,PhysRevC.103.064302}. %
We assume that the many-body wave function can be expressed as a superposition of quantum-number-projected basis configurations, given by 
\begin{equation}
  \vert\Psi^{J}_{NZ\sigma}\rangle = \sum_{K,q}f_{q,\sigma}^{JK}\vert J M K; N Z;q\rangle
\end{equation}
where $\vert J M K; N Z;q\rangle \equiv \hat{P}_{MK}^{J}\hat{P}^{N}\hat{P}^{Z}\vert\varphi(q)\rangle$. Here $\vert\varphi(q)\rangle$ represents a set of intrinsic states obtained by variation after particle-number projection (VAPNP)~\cite{ANGUIANO200262,PhysRevC.76.014308,PhysRevC.72.064303}, which is equivalent to determining the Bogoliubov states that are constrained to various amounts of expectation values $q$ for different collective operators and minimize the particle-number-projected energy. In this work we introduce a constraint on the expectation value of the quadrupole moment operator $\hat{Q}_{20}$ since we focus on the enhanced quadrupole collectivity in the island of inversion. $\hat{P}^{\prime}$s project intrinsic states onto a well-defined angular momentum $J$ and its $z$-component $M$, neutron number $N$, and proton number $Z$~\cite{Ring1980}. $f^{JK}_{q\sigma}$ is the weight function, where $\sigma$ is simply an enumeration index. It encodes the fluctuation of collectivity and can be given by solving the Hill-Wheeler equations~\cite{Ring1980}: 
\begin{equation}
    \sum_{K',q'} \left[\mathcal{H}_{KK^{\prime}}^{J}\left(q,q^{\prime}\right) - E^{J}_{\sigma}\mathcal{N}_{KK^{\prime}}^{J}\left(q,q^{\prime}\right)\right] f^{JK^{\prime}}_{q^{\prime},\sigma} = 0,
  \end{equation}
where the so-called Hamiltonian kernel $\mathcal{H}_{KK^{\prime}}^J(q; q^{\prime})$ and the norm kernel $\mathcal{N}_{KK^{\prime}}^J(q; q^{\prime})$ are obtained by
\begin{eqnarray}
\mathcal{H}_{KK^{\prime}}^{J}\left(q,q^{\prime}\right) &=&\langle\varphi\left(q\right)\vert H_{\mathrm{eff}}\hat{P}_{KK^{\prime}}^{J}\hat{P}^{N}\hat{P}^{Z}\vert\varphi\left(q^{\prime}\right)\rangle, \\
\mathcal{N}_{KK^{\prime}}^{J}\left(q;q^{\prime}\right) &=&  \langle\varphi\left(q\right)\vert\hat{P}_{KK^{\prime}}^{J}\hat{P}^{N}\hat{P}^{Z}\vert\varphi\left(q^{\prime}\right)\rangle.
\end{eqnarray}
With the PGCM many-body wave function at hand, one can subsequently compute electromagnetic moments and transition probabilities using consistently evolved operators including induced two-body parts. The formalism for consistently transforming transition operators within the IMSRG framework can be found in Ref.~\cite{PhysRevC.96.034324}. It has been applied to electromagnetic transitions~\cite{ZHOU2025139464,annurev:/content/journals/10.1146/annurev-nucl-101917-021120}, single-$\beta$ decays~\cite{GysbersNP2019}, and double-$\beta$ decays~\cite{PhysRevLett.124.232501}.

\section{Results and discussions}

A suitable valence space is of great importance for studying the island of inversion around the neutron number $N = 20$. 
To evaluate the effect from the extension of valence spaces, we derive effective Hamiltonians in three different valence spaces: the standard $sd$ space above an $^{16}$O core, the $sdf_{7/2}p_{3/2}$ space which includes both proton and neutron 0$f_{7/2}$ and 1$p_{3/2}$ orbits in addition to the \textit{sd} space, and the $sdfp$ space spanning two full major oscillator shells. For the underlying nuclear interaction, we adopt the 1.8/2.0 (EM) chiral interaction from Refs.~\cite{PhysRevC.83.031301,PhysRevC.93.011302,PhysRevC.96.014303}, which combines a free-space SRG-evolved NN interaction with an unevolved 3N force. This interaction accurately reproduces the ground-state energies of nuclei up to $A\approx100$~\cite{annurev:/content/journals/10.1146/annurev-nucl-101917-021120,PhysRevLett.120.152503}, with a root-mean-square deviation from the experiment of approximately 3.5 MeV across the light- and medium-mass regions~\cite{PhysRevLett.126.022501}. More recently, this interaction has been used to reproduce ground-state energies and low-lying states of tin isotopes beyond $^{132}$Sn within the VS-IMSRG(2)~\cite{PhysRevC.105.014302}.
The IMSRG evolution is carried out in a single particle basis with $e_{\mathrm{max}} = 12$, and 3N matrix elements are truncated at $E_{3\mathrm{max}} = 16$. As the primary application within a two-major shell valence space, we focus on the even-even Ne, Mg, and Si nuclei in vicinity of $N=20$ in this work. 

In multishell valence-space Hamiltonians, whether phenomenologically or microscopically constructed, the issue of potential center-of-mass (c.m.) contamination must be checked carefully. The Gl\"ockner-Lawson prescription~\cite{GLOECKNER1974313} is widely used to remove spurious excited states due to the c.m. motion. Here we exploit the procedure adopted in Ref.~\cite{PhysRevC.102.034320}, that is, add the Gl\"ockner-Lawson term at the beginning of the VS-IMSRG transformation. In this case, the initial Hamiltonian is written as $H = H_{\mathrm{intr}} + \beta H_{\mathrm{c.m.}} $, where \(H_{\mathrm{intr}}\) denotes the intrinsic Hamiltonian, and $ H_{\mathrm{c.m.}} = \mathbf{P}^2/2 A m + m A \tilde{\omega}^2 \mathbf{R}^2/2 - 3\hbar \tilde{\omega}/2 $ is the c.m. Hamiltonian. Note that \(\beta\) is a scaling parameter and \(\tilde{\omega}\) is the frequency of the c.m. Gaussian wave function. Such an initial Hamiltonian is then evolved via the VS-IMSRG transformation with the generator defined in Eqs.~(\ref{generator1})-(\ref{generator3}). It should be noted that the expectation value $\langle H_{\text{c.m.}}\rangle$ is close to zero if the c.m. wave function can be factorized by a single Gaussian~\cite{PhysRevC.82.034330}. The c.m. contamination has been shown to be adequately addressed through this procedure~\cite{PhysRevC.102.034320}. For further details, see Refs.~\cite{PhysRevC.96.034324,PhysRevC.82.034330}. However, as the valence space is enlarged, it is shown in some cases that the g.s. energies are shifted downward when the $\beta$ is large due to a substantial negative value of $\langle H_{\text{c.m.}}\rangle$~\cite{PhysRevC.102.034320}. The large negative value of $\langle H_{\text{c.m.}}\rangle$ implies that the induced many-body terms of $H_{\mathrm{c.m}}$ are large, and the IMSRG(2) approximation breaks down~\cite{PhysRevC.102.034320}. Therefore, to remove spurious c.m. modes but not to break the hierarchy of induced terms, we carefully adopt an appropriate $\beta$ value that keeps these induced terms under control by ensuring that the calculated expectation value of $\langle H_{\text{c.m.}}\rangle$ vanishes and the g.s energy remains consistent with the experimental data.

\begin{figure}[t]
  \begin{center}
    \includegraphics[width=\linewidth]{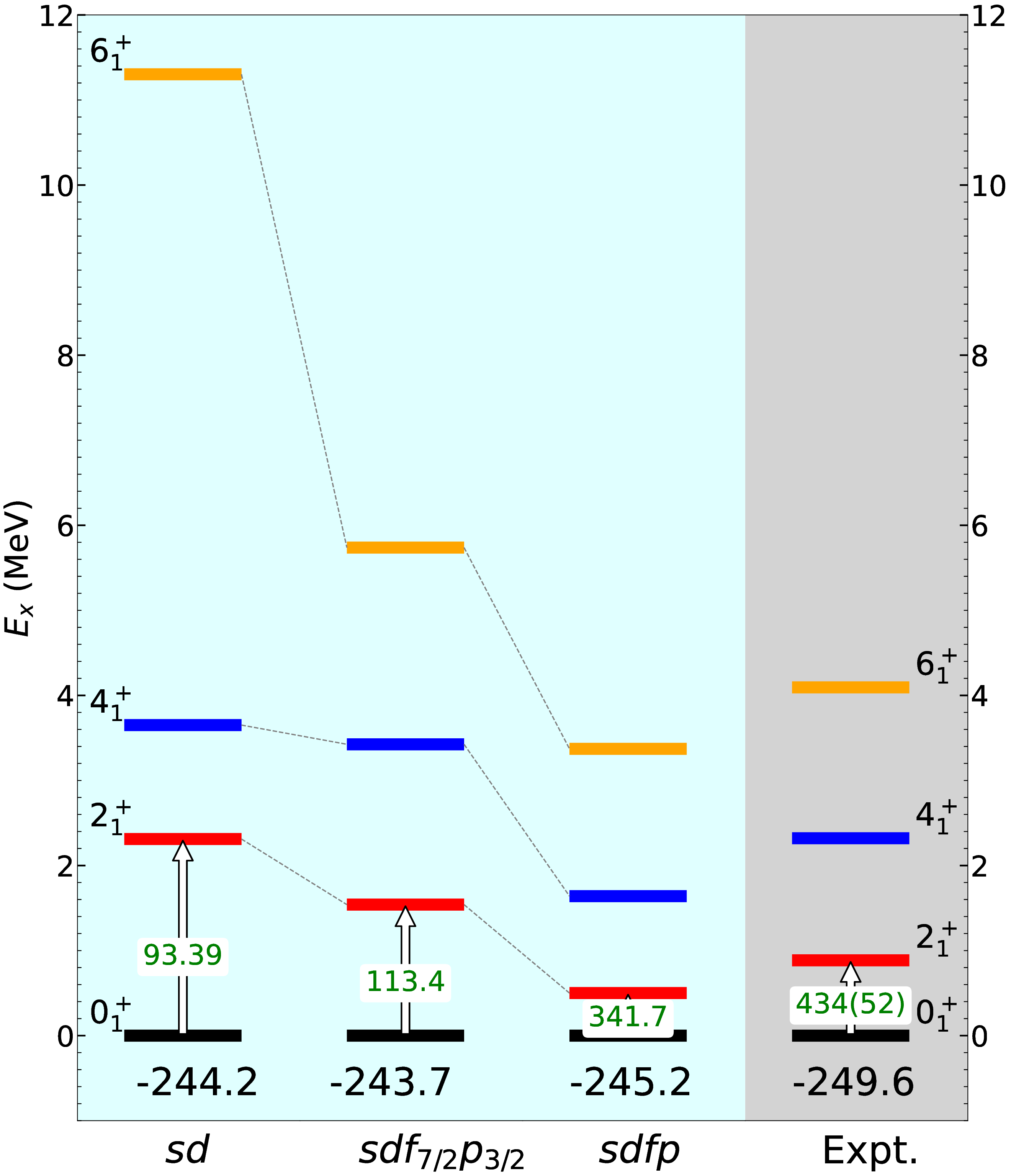}
    \caption{Calculated low-lying spectra of $^{32}$Mg obtained within the $sd$, $sdf_{7/2}p_{3/2}$, and $sdfp$ valence spaces, compared with experimental data. The arrows indicate the $B(E2)$ transition strengths from $0^{+}_{1}$ to $2^{+}_{1}$ in units of $e^2fm^4$. The experimental data are taken from the atomic mass evaluation (AME 2020)~\cite{Wang_2021} and the National Nuclear Data Center (NNDC)~\cite{NNDC}.}\label{Mg32 level in different model spaces}
  \end{center}
\end{figure}

\begin{figure}[t]
  \begin{center}
    \includegraphics[width=0.95\linewidth]{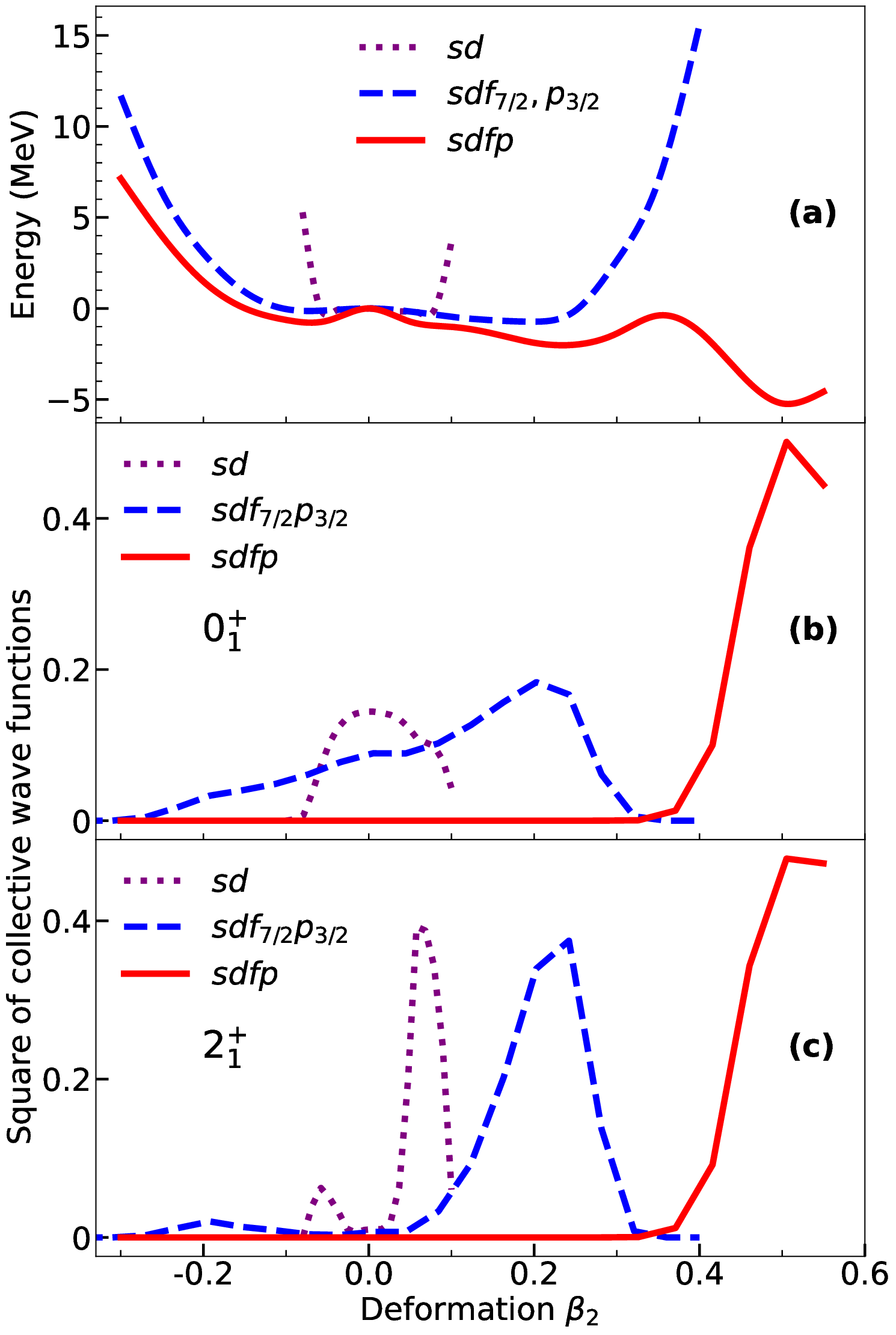}
    \caption{(a) Projected potential energy of $^{32}$Mg with particle-number projection and angular-momentum projection onto $J=0$ against axial deformation $\beta_2$, calculated in the $sd$, $sdf_{7/2}p_{3/2}$, and $sdfp$ spaces. For clarity, the spherical configuration energy has been set to zero. (b) The distributions of the collective wavefunction of the $0^{+}_{1}$ state in 3 different model spaces. (c) Same as panel (b) but for $2^{+}_{1}$ states.}\label{Mg32 PES after AMP}
  \end{center}
\end{figure}

In Fig.~\ref{Mg32 level in different model spaces}, we compare the low-lying spectra and the reduced $E2$ transition probabilities $B(E2;0_1^+\rightarrow2^+_1)$ of $^{32}$Mg calculated in the three model spaces with the experimental data. Note that the $E2$ transition operators are transformed consistently with the effective Hamiltonians by using the Magnus formulation of the IMSRG framework~\cite{PhysRevC.96.034324,annurev:/content/journals/10.1146/annurev-nucl-101917-021120}, and it is therefore unnecessary to introduce effective charges. The calculated g.s. energies are shown at the bottom, compared with the experimental data~\cite{Wang_2021}. Within the full \textit{sdfp} valence spaces, the g.s. energy decreases by a few MeV. We would expect a further lowering by a small amount from an exact diagonalization of the VS-IMSRG Hamiltonian. It should be mentioned that the exact diagonalizations of effective Hamiltonians within smaller valence spaces for the island of inversion are presented in Ref.~\cite{PhysRevC.102.034320}. Since we derive the \textit{sd}-shell effective Hamiltonian from the same 1.8/2.0 (EM) chiral interaction by means of the same IMSRG evolution, we can benchmark our PGCM calculations against the \textit{sd}-space results from Ref.~\cite{PhysRevC.102.034320}. For comparison, the exact diagonalization of \textit{sd}-shell effective Hamiltonian gives a g.s. energy of about $-245.7$ MeV for $^{32}$Mg (see Fig. 6 of Ref.~\cite{PhysRevC.102.034320}), which is slightly lower than our PGCM result. The excitation energy of the $2^+_1$ state and the $B(E2;0_1^+\rightarrow2^+_1)$ value given by our PGCM calculations are in great agreement with the ones obtained from shell-model calculations~\cite{PhysRevC.102.034320}, which justifies that the PGCM can provide a good approximation to the exact diagonlizations. %

Apparently, the calculated $2^+$ energy is overestimated and the $B(E2)$ value is underpredicted within the \textit{sd} valence space. This is consistent with the conclusion that the unexpected enhanced quadrupole collectivity is caused by intruder configurations from the \textit{fp} shell. As we expand the valence space, the excitation energies of the yrast $2_1^+$, $4_1^+$, and $6_1^+$ states are lowered. Within the full \textit{sdfp} shell, the calculated low-lying states exhibit an evident rotational behavior, in reasonably good agreement with the experimental data. The reduction of the $2^+$ energy implies an increase in quadrupole collectivity due to the intruder configurations from the \textit{fp} shell. 
Moreover, the reduced \textit{E}2 transition strength $B(E2; 0^+_1 \rightarrow 2^+_1)$ provides a direct probe of the quadrupole collectivity. As indicated by the arrows in Fig.~\ref{Mg32 level in different model spaces}, the calculated $B(E2; 0^+_1 \rightarrow 2^+_1)$ values increase substantially with the inclusion of more orbitals and tend to reproduce the adopted value~\cite{PRITYCHENKO20161}. The $B(E2; 0^+_1 \rightarrow 2^+_1)$ obtained within the full \textit{sdfp} space is 341.7 $e^2\text{fm}^4$, which is compatible with the shell-model result using an \textit{sdfp}-shell effective Hamiltonian derived from the EKK method and the effective charges $(e_p,e_n)=(1.25,0.25)e$~\cite{PhysRevC.95.021304}. Here we emphasize that our VS-IMSRG+PGCM calculations do not fit the single-particle energies or introduce the assumption of effective charges. Although the calculated $B(E2)$ is significantly improved compared to the VS-IMSRG calculations within smaller valence spaces, it is still a bit smaller than the adopted value. The VS-IMSRG has been found to systematically underestimate the $B(E2)$ values in well-deformed nuclei when truncated at the normal-ordered two-body level (NO2B)~\cite{PhysRevC.96.034324,HENDERSON2018468,PhysRevC.102.034320,PhysRevC.105.034333}.  
Moreover, we have attempted to further expand the valence space to include the 0$g_{9/2}$ shell. However, since the $sdfpg_{9/2}$ valence space is larger and more complex than the \textit{sdfp} space, we could not find an appropriate $\beta$ value to ensure that the expectation value of $\langle H_{\text{c.m.}}\rangle$ vanishes. It indicates that within the $sdfpg_{9/2}$ valence space, the center-of-mass treatment according to Ref.~\cite{PhysRevC.102.034320} spoiled the hierarchy of the induced many-body terms, resulting in a significantly over-bound ground-state energy. This presents a new challenge in the decoupling procedure of effective Hamiltonians within extremely large valence spaces.

\begin{figure}[t]
  \begin{center}
    \includegraphics[width=0.95\linewidth]{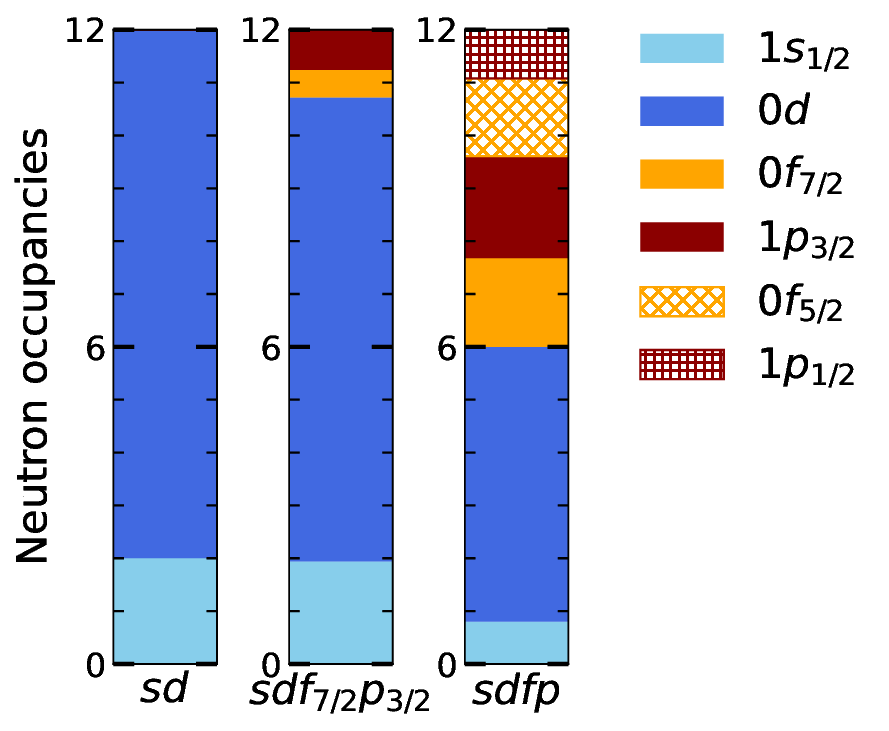}
    \caption{Neutron occupancies calculated within the $sd$, $sdf_{7/2}p_{3/2}$, and $sdfp$ spaces. Note that ``0\textit{d}'' denotes the occupation of the neutron $0d_{5/2}$ and $0d_{3/2}$ orbitals.}\label{Mg32 neutron occupancy}
\label{Mg32 model}
  \end{center}
\end{figure}

\begin{figure*}[t]
  \centering
  \makebox[\textwidth][c]{%
  \includegraphics[width=\textwidth]{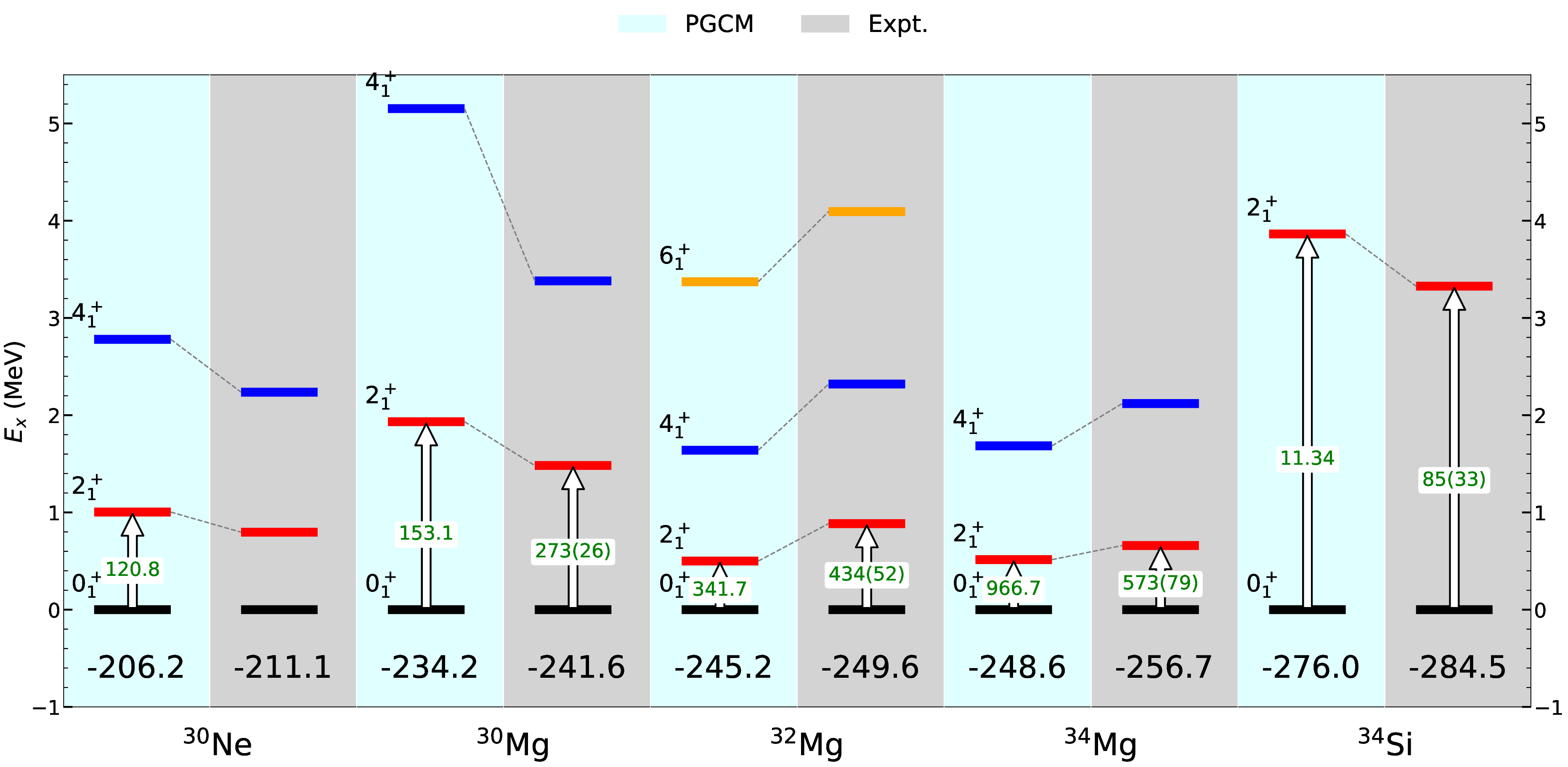}
}
  \caption{Calculated low-lying spectra of $^{30}$Ne, $^{30-34}$Mg, and $^{34}$Si within $sdfp$ model space, compared with experimental data. The arrows indicate the $B(E2;0^+_1\rightarrow2^+_1)$ values in units of $e^2fm^4$. The experimental data are taken from the AME 2020~\cite{Wang_2021} and NNDC~\cite{NNDC}. The calculations are done with $e_{max}=12$ and $E_{3\mathrm{max}}=16$.}
  \label{different nuclei levels}
\end{figure*}

It is intriguing to see the evolution of nuclear quadrupole deformation with the gradual expansion of the valence space and its impact on the low-lying structure of $^{32}$Mg. %
Figure~\ref{Mg32 PES after AMP}(a) illustrates the projected potential energy of $^{32}$Mg against the axial quadrupole deformation ${\beta_2}$ calculated within the three different valence spaces, all with projection of particle number onto $Z=12$ and $N=20$, and angular momentum onto $J = 0$. %
As the model spaces expand, the projected potential energy curves (PECs) span a wider range of nuclear quadrupole deformation. This is mainly because the limited valence space lacks the necessary single-particle degrees of freedom, in particular high-$j$ orbitals. The substantial energy splitting of these orbitals along with the increase of deformations, as a result of strong quadrupole-quadrupole correlations, is essential for achieving the energy gain required to stabilize a large deformation. The minimum shifts from near-sphericity within the $sd$ space to a large prolate deformation of $\beta_2\approx0.5$ within the \textit{sdfp} space. Note that our \textit{sdfp}-shell projected PEC is strikingly similar to the projected PEC given by IM-GCM employing a Hamiltonian appropriately evolved by MR-IMSRG~\cite{ZHOU2025139464}, although the limited valence space in our calculations prevents constraining to larger deformations of $\beta_2\geqslant 0.6$. 

In addition, we show the distribution of collective wave functions for the $0^+$ ground state and the $2_1^+$ excited state computed within three valence spaces in Fig.~\ref{Mg32 PES after AMP}(b) and Fig.~\ref{Mg32 PES after AMP}(c), respectively. %
The distribution of collective wave functions is defined to account for the probability density, normalized to 1, of finding the state $\vert\Psi^J_{NZ\sigma}\rangle$ with given deformation parameters $\beta_2$ (see more details in Refs.~\cite{PhysRevC.81.044311,PhysRevC.81.064323}). %
One can see that the $0_1^{+}$ and $2_1^{+}$ states obtained within the \textit{sd} space are dominated by near-spherical configurations with $\vert\beta_2\vert\leqslant0.08$. %
In contrast, the distribution of the $0_1^{+}$ state within the $sdf_{7/2}p_{3/2}$ space exhibits a remarkable mixture of moderately prolate and oblate deformed configurations in the region with $\vert\beta_2\vert\leqslant0.3$, which leads to two shallow minima for prolate and oblate deformation in the projected PEC, respectively. 
Finally, in the $sdfp$ model space, the $0_1^{+}$ and $2_1^{+}$ states are dominated by strongly prolate deformed configurations with $\beta_2\approx0.5$, indicating a significantly more deformed state. %
The distribution of collective wave functions clearly demonstrates that dominant configurations gradually shift towards stronger quadrupole collectivity along with the expansion of the valence space, thereby highlighting the necessity of employing a two-major-shell valence space to fully capture the enhanced collectivity of the island of inversion. 

Previous studies suggested that the onset of enhanced quadrupole collectivity is due to  \textit{mp-mh} excitations over the $N=20$ shell gap for the ground state. Fig~\ref{Mg32 neutron occupancy} depicts the neutron occupancies obtained with the VS-IMSRG Hamiltonians decoupled to three valence spaces. It demonstrates that, as we gradually expand the valence spaces, more neutrons are excited from the \textit{sd} to the \textit{fp} shell. Note that our VS-IMSRG+PGCM result within the $sdf_{7/2}p_{3/2}$ space shows that $\sim 1.5$ neutrons occupy the \textit{fp} shell, which is compatible with the previous $\pi$\textit{sd}+$\nu$\textit{sd}$f_{7/2}p_{3/2}$ shell VS-IMSRG calculation~\cite{PhysRevC.102.034320}. Within the full \textit{sdfp} space, more neutrons are promoted to the \textit{fp} shell. Inclusion of neutron $0f_{5/2}$ and $1p_{1/2}$ orbitals not only allows neutrons to excite and occupy these two orbitals, but also enhances the occupation of the $0f_{7/2}$ and $1p_{3/2}$ orbitals. It may imply that the PGCM calculation within a larger valence space captures more collective correlations associated with multi-particle multi-hole (\textit{mp-mh}) excitations. Fig.~\ref{Mg32 neutron occupancy}, combined with Fig.~\ref{Mg32 level in different model spaces}, indicates the correlation between the onset of enhanced quadrupole collectivity and the cross-shell \textit{mp-mh} excitations, as well as the importance of larger valence spaces.   

\begin{table}[htbp]
\setlength{\tabcolsep}{0pt}
\renewcommand\arraystretch{1.2}
\setlength{\arrayrulewidth}{0.3mm} 
    \centering
    \caption{Convergence of the ground-state energies (in MeV) of $^{30}$Ne, $^{32}$Mg and $^{34}$Si with different $e_{\text{max}}$.}
\begin{tabular*}{\columnwidth}{@{\extracolsep{\fill}}lcccc}
        \hline \hline
       Nuclei & $e_{\text{max}}=6$ & $e_{\text{max}}=8$ & $e_{\text{max}}=10$ &$e_{\text{max}}=12$  \\ 
        \hline
        $^{30}$Ne & -189.7 & -200.5   & -204.6 &-206.2 \\
        $^{32}$Mg & -224.7 & -236.9   & -241.5 &-245.2 \\
        $^{34}$Si & -256.2 & -270.2   & -275.5 &-276.0 \\
        \hline \hline
    \end{tabular*}
    \label{convergence of emax}
\end{table}

We broaden our calculations to investigate the low-lying states of $^{30,34}$Mg and the $N=20$ isotones $^{30}$Ne and $^{34}$Si, using the \textit{ab initio} $sdfp$-shell Hamiltonians derived from VS-IMSRG. We first check the convergence with different $e_\mathrm{max}$ values and list the calculated ground-state energies in Table~\ref{convergence of emax}. It is evident that the ground-state energy has already converged for $^{34}$Si at $e_\mathrm{max}=12$, while the energies of $^{30}$Ne and $^{32}$Mg obtained with $e_\mathrm{max}=12$ still decrease by approximately 1.6 and 3.7 MeV compared to those obtained with $e_\mathrm{max}=10$, respectively. The notably smaller decrease when moving towards $e_\mathrm{max}=12$ indicates that the calculations are approaching convergence. The calculated g.s. energies, level spectra and $B(E2;0_1^+\rightarrow2^+_1)$ values are compared with the experimental data in Fig.~\ref{different nuclei levels}. %
For all investigated nuclei, our calculations reproduce the observations reasonably well, whereas the previous VS-IMSRG calculations within smaller model spaces systematically overestimated the $2^+_1$ excitation energies and underestimated the $B(E2;0_1^+\rightarrow2^+_1)$ values~\cite{PhysRevC.102.034320}. As the neutron numbers increase in the Mg isotopes, the calculated excitation energies of the $2_1^+$ states show a notable lowering at $N=20$. Although the calculations exaggerate the lowering in a certain context, it is in accordance with the experimental trend. Also, the $B(E2)$ values increase along with the neutron numbers around $N=20$ for Mg isotopes, exhibiting the disappearance of the $N=20$ shell gap and the formation of the island of inversion.

For $N=20$ isotones, our \textit{sdfp}-shell calculations give a moderate lowering of the $2^+_1$ energy and a slight enhancement of the $B(E2; 0^+_1 \rightarrow 2^+_1)$ value for $^{30}$Ne compared to the previous $\pi$\textit{sd}+$\nu$\textit{sd}$f_{7/2}p_{3/2}$ shell VS-IMSRG calculations~\cite{PhysRevC.102.034320}, improving the agreement. In contrast, the calculated $2^+_1$ energy of $^{34}$Si is much higher than that of $^{30}$Ne and $^{32}$Mg, and the $B(E2)$ value is significantly lower. It indicates the dominant role of the spherical configuration in the $0_1^+$ ground state and the $2^+_1$ excited state of $^{34}$Si. Note that our \textit{sdfp}-shell calculations reproduce well the observed $2^+_1$ energy of $^{34}$Si, but give a $B(E2;0^+_1\rightarrow2^+_1)$ value that is smaller than the adopted value. This may be attributed to the fact that the \textit{ab initio} methods based on spherical references tend to underestimate $B(E2)$ values due to missing contributions from \textit{mp-mh} excitations~\cite{PhysRevC.102.034320,PhysRevC.96.034324,HENDERSON2018468}. Combining $2_1^+$ energies and $B(E2)$, we demonstrate an apparent collapse of the $N=20$ shell closure in $^{30}$Ne and $^{32}$Mg and its persistence in $^{34}$Si. Overall, we found that the VS-IMSRG+PGCM within the full two-major-shell valence space provides a reasonably good description of the border of the island of inversion.

\begin{figure}[t]
  \begin{center}
    \includegraphics[width=\linewidth]{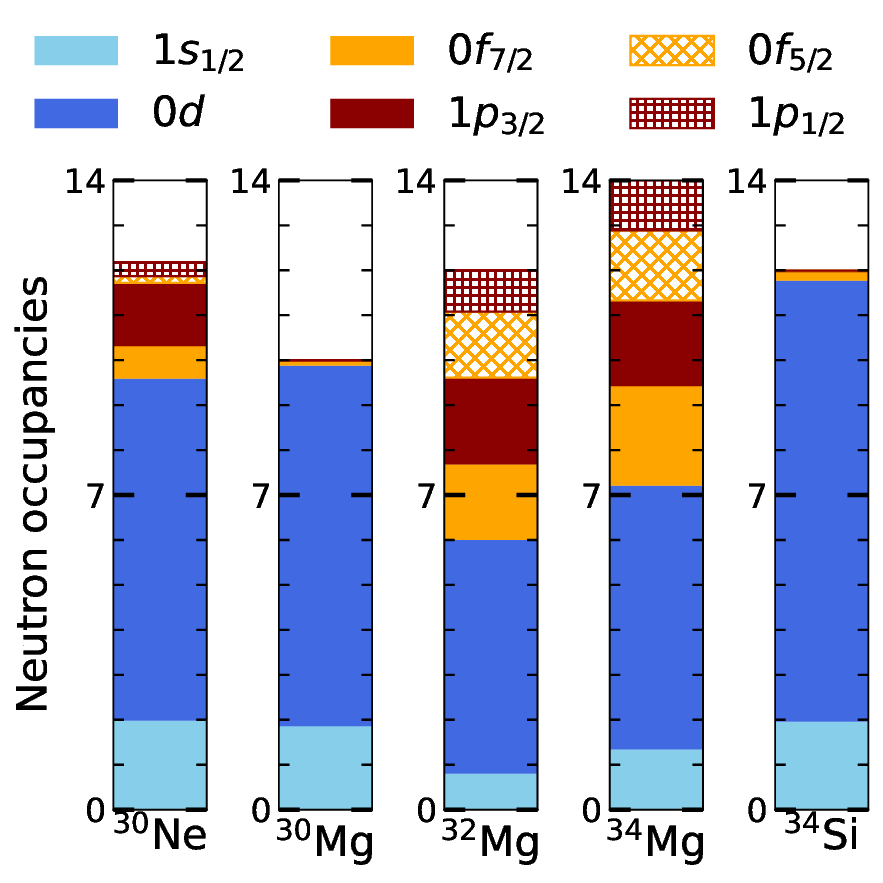}
    \caption{Neutron occupancies calculated within the full $sdfp$ space for the ground states of $^{30-34}$Mg and the adjacent $N=20$ isotones $^{30}$Ne, $^{34}$Si. Note that ``0\textit{d}'' denotes the occupation of the neutron $0d_{5/2}$ and $0d_{3/2}$ orbitals.}
\label{diff model}
  \end{center}
\end{figure}

The neutron occupancies computed within the full  \textit{sdfp} space for the ground states of $^{30}$Ne, $^{30-34}$Mg, and $^{34}$Si are summarized in Fig.~\ref{diff model}, illustrating the evolution of cross-shell \textit{mp-mh} excitations along the isotopic and isotonic chains. For $^{30-34}$Mg, we see a sudden jump in neutron occupancies of the \textit{fp} shell at $N=20$. It is consistent with earlier phenomenological shell model studies~\cite{PhysRevC.60.054315,PhysRevC.90.014302}, the empirical analysis given by a three-level mixing model~\cite{PhysRevC.94.051303}, and the VS-IMSRG study within smaller valence spaces~\cite{PhysRevC.102.034320}. A further moderate increase in neutron occupancies of the \textit{fp} shell is presented for $^{34}$Mg.   
For $N=20$ isotones, about 2 neutrons are excited to the $fp$ shell in the ground state of $^{30}$Ne. In contrast, the ground state of $^{34}$Si is shown to be dominated by neutrons occupying the \textit{sd} shell, clearly indicating the persistence of the $N=20$ shell closure. We note that our calculation overpredicts the \textit{fp}-shell occupancies for $^{32,34}$Mg when compared to the conventional picture for $^{32}$Mg being basically a 2p-2h state, as suggested by phenomenological shell-model studies~\cite{PhysRevC.90.014302,PhysRevC.60.054315,FUKUNISHI1992279}. Note that the semi-microscopic MBPT approach using the EKK method~\cite{PhysRevC.95.021304} and the empirical three-level mixing model~\cite{PhysRevC.94.051303} also suggested a large probability of higher \textit{mp-mh} configurations in the $0_1^+$ state of $^{32}$Mg. Also, the occupation numbers are model-dependent, and so calculations
employing different Hamiltonians need not agree on them. Therefore, no strong conclusions should be drawn here.

Finally, it is of particular interest to compare the low-lying states given by our VS-IMSRG+PGCM framework with that obtained from the IM-GCM calculations~\cite{ZHOU2025139464} which combines the PGCM with the MR-IMSRG rather than the VS-IMSRG. We note the striking similarity between the projected PEC's obtained with these two methods. Both approaches consistently reproduce the observed onset of large quadrupole collectivity from $^{30}$Mg to $^{34}$Mg indicated by the compression of the low-lying rotational bands and the increase in $B(E2; 0^+_1 \rightarrow 2^+_1)$. Our VS-IMSRG+PGCM calculations present slightly compressed level spectra for $^{32,34}$Mg when compared to the experimental data, whereas the IM-GCM ones are somewhat stretched. Moreover, the IM-GCM quantitatively reproduces the adopted $B(E2; 0^+_1 \rightarrow 2^+_1)$ values within the experimental errors, while our VS-IMSRG+PGCM calculations still slightly underpredict the $B(E2; 0^+_1 \rightarrow 2^+_1)$ values for $^{30}$Mg and $^{32}$Mg. This underestimation of the electromagnetic transition strengths, despite a remarkable improvement through expanding the valence spaces, may be attributed to normal ordering with respect to an ensemble of spherical reference states rather than an ensemble of multiple symmetry-restored Hartree-Fock-Bogoliubov states that are used in the IM-GCM~\cite{ZHOU2025139464}. 

\section{Summary and conclusions}
In this work, we have used the VS-IMSRG framework to construct \textit{ab initio} multishell valence space Hamiltonians, starting from chiral \textit{NN} and 3\textit{N} forces. To handle the effective Hamiltonians within large valence spaces, we employed the PGCM to variationally approximate the conventional shell model. Without further empirical fitting or the introduction of effective charges, we study the nuclear spectroscopy of strongly deformed open-shell nuclei in the island of inversion around $N=20$. The expanded model space markedly improves the description of collective properties, e.g., excitation energies of $2^+$ states and $E2$ transition strengths in nuclei inside or outside of the island of inversion, confirming the disappearance of the $N=20$ shell closure and the onset of intruder configurations driven by cross-shell \textit{mp-mh} correlations. The calculated neutron occupancies, which reflect intruder configurations arising from the \textit{mp-mh} excitations across the $N=20$ shell gap, are consistent with the conclusions of the phenomenological studies.  

The reasonably good description of the island of inversion reveals that, within suitable multishell valence spaces, the VS-IMSRG+PGCM approach is capable of capturing collective correlations associated with cross-shell \textit{ph} excitations, and accurately describing the onset of enhanced quardupole collectivity due to dominant intruder configurations. Since the VS-IMSRG+PGCM approach allows us to add or remove valence orbits and collective coordinates, we can elucidate the essential degrees of freedom for describing a many-body wave function. While we need to carefully check the center-of-mass contamination when we choose an appropriate multishell valence space, this method provides us with an opportunity for \textit{ab initio} valence-space-based studies with tractable computational cost on the heavier open-shell nuclei, in which nucleons are strongly correlated.

\section*{Acknowledgements}
The authors would like to thank J. M. Yao for fruitful discussions and useful comments. We thank T. Miyagi for the \texttt{NuHamil} code~\cite{Miyagi2023} which was used to generate chiral EFT matrix elements, and R. Stroberg for \texttt{IMSRG++}~\cite{Stroberg_IMSRG} which was used to perform the VS-IMSRG decoupling. The authors also thank B. Bally and T. R. Rodr\'{i}guez since the PGCM code was developed based on \texttt{TAURUS}~\cite{Bally2021,Bally2024}. This material is based on the work supported by the National Natural Science Foundation of China under Grant No. 12275369.  




\bibliographystyle{elsarticle-num} 
\bibliography{IoI}






\end{document}